\documentclass[10pt, a4paper]{article}
\usepackage{lrec2016}
\usepackage{multibib}
\newcites{languageresource}{Language Resources}
\usepackage{graphicx}

\usepackage{epstopdf}
\usepackage[latin1]{inputenc}

\title{RankDCG: Rank-Ordering Evaluation Measure\\ 
}

\name{Denys Katerenchuk, Andrew Rosenberg}

\address{ CUNY Graduate Center, New York, USA \\
365 Fifth Avenue, Room 4319, New York, NY 10016\\
CUNY Queens College, New York, USA \\
65-30 Kissena Boulevard, Room A-202, Flushing, NY 11367\\
dkaterenchuk@gradcenter.cuny.edu, andrew@cs.qc.cuny.edu\\}

\abstract{
Ranking is used for a wide array of problems, most notably information retrieval (search). There are a number of popular approaches to the evaluation of ranking such as Kendall's $\tau$, Average Precision, and nDCG. When dealing with problems such as user ranking or recommendation systems, all these measures suffer from various problems, including an inability to deal with elements of the same rank, inconsistent and ambiguous lower bound scores, and an inappropriate cost function. We propose a new measure, rankDCG, that addresses these problems. This is a modification of the popular nDCG algorithm. We provide a number of criteria for any effective ranking algorithm and show that only rankDCG satisfies all of them. Results are presented on constructed and real data sets. We release a publicly available rankDCG evaluation package. \\ 
\newline 
}

\begin{document}

\maketitleabstract

\section{Introduction}


Every algorithm needs to be assessed to determine its performance. No single measure can be applied to all problems. 
If we consider a single area of computer science natural language processing (NLP), each problem requires a specific evaluation method. 
For example, for a simple classification task, accuracy is an intuitive and useful measure \cite{Dumais:1998:ILA:288627.288651,katerenchuk2014your}. For named entity recognition and other ``detection'' tasks with a relatively small percent of relevant items \cite{TjongKimSang:2003:ICS:1119176.1119195,katerenchuk2014improving}, F-measure \cite{Rijsbergen:1979:IR:539927} is best suited. Correlation measures such as Pearson's r \cite{pearson1895note} and Spearman's $\rho$ \cite{spearman1904} are used to find relationships between entities \cite{strapparava2008learning,schullerinterspeech}. 
Kendall's $\tau$ \cite{kendall38}, Average Precision (AP) \cite{zhuap}, Mean Average Precision (MAP) and Discounted Cumulative Gain (DCG) \cite{Jarvelin:2002:CGE:582415.582418} are all used in information retrieval (IR) and ranking type of problems \cite{lapata2006automatic,philbin2007object,jarvelin2000ir}.

Despite a large number of different ranking measures, there are still problems that cannot be appropriately evaluated. In particular, when the task is to rank discrete value elements with multiple ties of the same rank and a skewed rank distribution. This type of problem often arises in a number of ranking problems such as information retrieval or search. While some measures address parts of this problem, none addresses all of them.

In this paper, we propose a new measure to deal with rank-ordering problems. We start with defining the problem and criteria that need to be satisfied in section \ref{problem} In Section 3 we give a short survey of available evaluation measures. In Section \ref{analysis} we propose rankDCG, an improved evaluation measure and evaluate its performance in Section \ref{experiments} We sum up our findings and conclude our work in Section \ref{conclusion}

\section{Ordering}\label{problem}
The problem of ordering is well known. It involves tasks such as internet search. The objectives are to find and order information from a near infinite set of data, namely web pages. Formally it can be defined as follows:

\begin{quote}
Given a list of elements A = $[ x_1, x_2, x_3, ..., x_n]$, objective is to find list B = $[ x | x \in A, f(x) > 0] $, where $f(x)$ is a relevance function that returns a rank that is higher or equal to 0. Often additional objective is applied such as B = $[f(x_1) > f(x_2) > f(x_3) > ... > f(x_m)]$ where m is a number of relevant document with $f(x) > 0$ and $m \leq n$.
\end{quote}

In order to evaluate this problem, a comparison between two lists, a reference and a hypothesis, is needed. Relevance and ordering are the two prime factors that need to be considered. Because most measures were designed for IR tasks, the relevance of elements plays a crucial role in determining the evaluation score lower bound. In other words, if all elements in the hypothesis list are irrelevant, the score should be equal to 0 or some other lower bound. \\


In this paper, we consider an ordering problem that often appears in real word problems such as recommendation systems and user ranking. These tasks might appear identical to the web search problem. However, there is a number of distinct characteristics. 
First of all, each element is relevant (no irrelevant entities). Second, the element ranks are discrete values. Third, the rank values are not unique. In other words, there are many elements of the same rank (multiple ties). Lastly, the elements might not follow the normal distribution of rank values. This case is also common to web search where only very few top results are relevant and the majority are somewhat related or not relevant to the query at all. 
Formally the problem can be described in the following way:

\begin{quote}
Given a list of elements A = $[ x_1, x_2, x_3, ..., x_n]$, 
objective is to find list B = $[f(x_1) \geq f(x_2) \geq f(x_3) \geq ... \geq f(x_n)]$ where $f(x)$ is a rank function that returns rank $r$ 
with $r \in$ \math{N}$ and n is a number of elements.
\end{quote}

All conventional evaluation measures have a number of shortcomings while evaluating this type of problem. For this reason, we propose a set of criteria that an evaluation measure. This measure needs to address the following objectives: 
\begin{enumerate}
  \item correctly work with multiple ties
  \item address non-normal rank value distribution
  \item emphasize correct ordering of high rank elements
  \item produce consistent and meaningful scoring range
\end{enumerate}

In the next section, we will survey available algorithms and highlight some drawbacks of the most common rank evaluation measures.

\section{Evaluation Measures Survey}\label{overview}
Multiple rank-ordering evaluation metric algorithms exist in the field of information retrieval (IR). However, none of them is appropriate for the task described in the previous chapter. Keeping in mind the specifics of the problem, we survey various metrics, analyze their performance, and underline drawbacks.

\subsection{F-measure}

F-measure or F-score \cite{Rijsbergen:1979:IR:539927} is a common evaluation measure that is used to measure IE algorithms such as search \cite{Peng:2006:IER:1142097.1142104}. This measure is defined as follows:
\begin{center}
$$F = 2*\frac{p * r}{p + r},$$\\ 
\end{center}
where p - precision and r - recall\\

F-measure takes into account precision and recall. Precision measures the portion of retrieved elements that are relevant. Recall measures the portion of relevant elements that were discovered.  However, this measure is not appropriate for problems where all elements are relevant.  In addition, this measure does not take into consideration different ranks. F-measure only evaluates the number of relevant elements. Therefore, it is not suitable for a rank-ordering evaluation. 

\subsection{Average Precision and Mean Average Precision}

Average Precision (AP) \cite{zhuap} is a measure that is designed to evaluate IR algorithms. AP can deal with non-normal rank distribution, where the number of elements of some rank is dominant. AP measures precision at each element, multiplies the change in recall from the previous step, and averages over all list elements.  There exists a variation of AP that takes into consideration only the first k elements. However, since we are concerned with a ranking of all elements, we will not focus on this variant. The formula to calculate the AP is the following:
\begin{center}
$$AP = \sum\limits_{k=1}^n P(k) * \Delta R(k)$$
\end{center}
where $P(k)$ = precision@$k$ and $\Delta R(k) = |recall(k-1) - recall(k)|$.\\

Researchers often use mean average precision (MAP) \cite{liu_mean_2009}, which is defined as the mean of AP over multiple information retrieval lists.
\begin{center}
 $$MAP = \frac{\sum\limits_{q\in Q} AP(q)}{|Q|},$$
\end{center}
where Q = a set of ordering problems and q = a single evaluation instance. \\

Both AP and MAP measures have been designed to evaluate rank-ordering problems. The measures, however, assume no ties among ranks which manifests in inconsistent lower bounds.  Furthermore, these measures evaluate all rank values with equal cost. However, the problem described in Section 2 requires more emphasis on ordering of rare high-rank elements and less for low-rank elements since these elements are not as important and often over-represented. This creates a problem where misplacing a low-rank element can produce a low score, despite the fact that this element might not be very relevant to an otherwise a good ordering result. More detail of this case can be found in Section \ref{experiments}

\subsection{Kendall's $\tau$}

Kendall's $\tau$ \cite{kendall38} is a correlation measure. This measure is often used when evaluating rank-ordering results. The measure considers the number of element pairs in reference and hypothesis lists and checks whether the element positions correlate. The formal definition of Kendall's $\tau$ is shown below:

\begin{center}
$$\tau = \frac{c - d}{\frac{1}{2} n(n-1)},$$ 
\end{center}
where c - a number of concordant (i.e. a correct relative ranking) pairs and d - a number of discordant (i.e. an incorrect relative ranking) pairs.\\

Kendall's $\tau$ is a popular choice for rank evaluation. Unfortunately, this measure also has some drawbacks. First of all, it does not explicitly deal with multiple ties and non-normal rank distribution. This will lead to a problem when an algorithm assigns the same (majority) rank value to all elements. Secondly, Kendall's $\tau$ does not produce a consistent lower bound score when the ranks follow a non-normal distribution. In addition, the score is produced by comparing the number of correlated elements and it does not emphasize rare high-rank elements. For these reasons, Kendall's $\tau$ is not the best choice to evaluate rank-ordering problems defined in Section \ref{problem}


\subsection{Discounted Cumulative Gain}

Among all evaluation measures, Discounted Cumulative Gain (DCG) \cite{Jarvelin:2002:CGE:582415.582418} has multiple advantageous characteristics to address a rank-ordering problem mentioned in the previous section. For this reason, it is often used in research  \cite{lapata2006automatic,philbin2007object,jarvelin2000ir}. The main distinction of DCG from others measures is the ability to address non-normal rank distribution by assigning a higher cost to high-rank elements. This emphasizes the high-rank element identification. The formal definition of DCG is defined below:

\begin{center}
$$DCG = \sum_{i=1}^n \frac{rel(x_i)}{log_2(i+1)},$$\\
\end{center}
where n - a number of elements and rel() - some relevance function of the i-th element in a given list.\\

For comparison across multiple tasks, a normalized variant of DCG, nDCG, is calculated in the following way:

\begin{center}
$$nDCG=\frac{DCG}{IDCG},$$ 
\end{center}
where IDCG - represents the ideal DCG.\\

This evaluation also has drawbacks. The first drawback is this evaluation metric was designed for information retrieval rather than ordering evaluation. This means that this measure considers the number of relevant and irrelevant documents. Since all elements in the rank-ordering task defined in Section \ref{problem} are relevant, the measure's lower bound is never equaled to zero. As a result, the range of prediction is from 1 in the best case ordering to some arbitrary number between 1 and 0. This factor makes results hard to understand because an nDCG score of 0.56 might be the worst case ordering.
Another drawback is the cost function puts too much stress on the high-rank element identification. The cost function was designed in this way intentionally to bring more relevant search results closely to the top. However, the rank-ordering problem needs a relative function with respect to the rest of the elements. Lastly, standard DCG produces different cost based on the element positioning. For example a list [9,1,1] will have different costs for [1,9,1] and [1,1,9]. However, we contend that the two lists are equally wrong because the algorithm decided that element of rank 9 is rank 1. The permutations inside rank subgroup should not matter in the evaluation process. 

\begin{figure*}
\centering
\begin{minipage}{.42\textwidth}
  \includegraphics[width=1\linewidth]{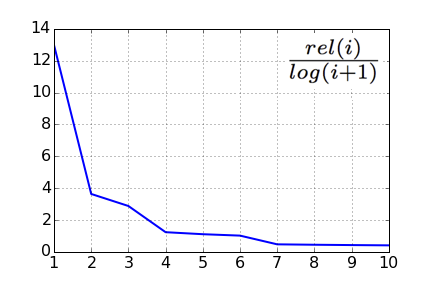}
  \caption{Cost function: $\frac{rel(i)}{log(i+1)}$}
  \label{pict1}
\end{minipage}%
\begin{minipage}{.42\textwidth}
  \includegraphics[width=1\linewidth]{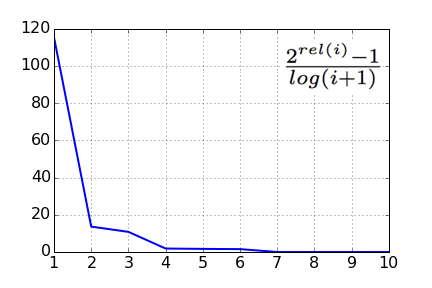}
  \caption{Cost function: $\frac{2^{rel(i)}-1}{log(i+1)}$}
  \label{pict2}
\end{minipage}
\begin{minipage}{.42\textwidth}
  \centering
  \includegraphics[width=1\linewidth]{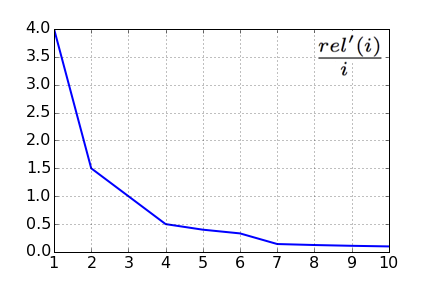}
  \caption{Cost function: $\frac{rel'(i)}{i}$}
  \label{pict3}
\end{minipage}%
\begin{minipage}{.42\textwidth}
  \centering
  \includegraphics[width=1\linewidth]{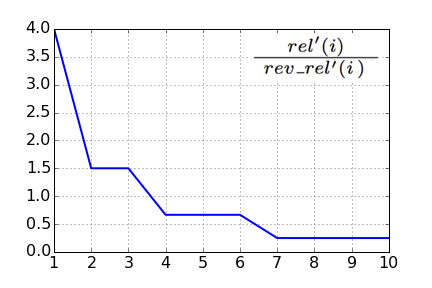}
  \caption{Cost function: $\frac{rel'(i)}{rev\_rel'(i)}$}
  \label{pict4}
\end{minipage}
\end{figure*}

\section{Rank Discounted Cumulative Gain}\label{analysis} \label{rankDCG}

In this section, we present a novel measure which we call rank discounted cumulative gain (rankDCG). This measure is a modified version of the popular nDCG algorithms. From Section \ref{overview}, we can see that conventional evaluation measures fall short from addressing evaluation criteria. In particular, a good measure for rank-ordering problems needs to address the following:

\begin{enumerate}
  \item multiple ties
  \item non-normal rank value distribution
  \item emphasis on high-rank elements
  \item consistent scoring range
\end{enumerate}

In order to demonstrate our algorithm, we start with constructing an example problem. The list L that is shown below, is ordered by rank values. In other words, each element represents an output from $rel(i)$, a relevance function. The rank values are discrete and the list contains multiple ties of elements with the same rank. In addition, the element rank distribution is non-normal.

\begin{center}
$L = [9_1,4_2,4_3,2_4,2_5,2_6,1_7,1_8,1_9,1_{10}]$\\
\end{center}

The first property rankDCG needs to address is non-normal rank distribution. From the Section \ref{overview}, we saw that most rank measures, with the exception of DCG, do not have a way to distinguish between low-rank and high-rank elements in the scoring function. For this reason, we consider a number of cost functions that are similar to the DCG definition. We experiment with four different functions performed on list L and plot them in figures \ref{pict1}-\ref{pict4}. The x-axis of each plot is the element order in list L and the y-axis is a cost generated from the experimental functions.\\

We start with an analysis of two cost functions: the standard DCG cost function and a modified version used in \cite{Burges:2005}. From the Figures \ref{pict1} and \ref{pict2}, we can see that both functions put more than half of their weight on the correct identification of the highest element. This can introduce bias toward finding the top-rank element rather than ordering. To address this issue, we design a function $rel'(i)$ that produces an element rank based on the number of unique element ranks in the list. The list L contains ten relative rank values, but only four unique values. We create a mapping function to assign a unique rank based on the rank subgroup. In other words, the top-rank element is equal to the size of the element rank set, $|\{L\}|$. Every following distinct element-rank will have its rank decreased by one. The results are plotted in Figure \ref{pict3}. In this case, given list $L$ to the function $rel'(i)$ we get a corresponding list $L'$ with the following ranks:\\

\begin{center}
$L' = [4_1,3_2,3_3,2_4,2_5,2_6,1_7,1_8,1_9,1_{10}]$\\
\end{center}

In addition to the above modification, we modify the discounting factor in the denominator of the DGC formula. DCG's discounted factor relies on the position of each element, and this implies that the last four value of $L'$ list will produce different costs. Instead of using the elements' position, we find that reversed mapping order of $rel'()$ function works the best for discounted factor. The mapping between elements in $L'$ and the discounted factors are represented in list $D$ and the final cost function is shown in Figure \ref{pict4}. This discounted factor creates a step-wise function that eliminates a chance of getting a different score from permutations inside element subgroups.\\ 

\begin{center}
$D = [1_1,2_2,2_3,3_4,3_5,3_6,4_7,4_8,4_9,4_{10}]$\\
\end{center}

At this point the cost function is the following:

\begin{center}
$$DCG' = \sum_{i=1}^n \frac{rel'(i)}{rev\_rel'(i)},$$\\
\end{center}
where n - a number of elements, $rel'(i)$ - cost function that takes L and creates L' and $rev\_rel'(i)$ - reversed rel'(i) function that takes L and creates discounted factor for each element that is shown in list D.\\ 

At last, we normalize $DCG'$ to create a meaningful and consistent lower bound. The final normalized version of rankDGC is defined below:

\begin{center}
$$rankDCG=\frac{DCG' - min(DCG')}{max(DCG') - min(DCG')}$$
\end{center}

Python implementation of rankDCG is available for download at our website (http://speech.cs.qc.cuny.edu).  

\begin{table*}
\centering
\begin{tabular}{ | l | c | c | c | c | c |}
\hline
\# & Hypothesis List & Kendall's $\tau$ & AveP & nDCG & rankDCG \\  
\hline\hline
1 & [9, 4, 4, 2, 2, 2, 1, 1, 1, 1] & 1.0 & 1.0 & 1.0 & 1.0 \\
2 & [9, 4, 4, 2, 2, 1, 2, 1, 1, 1] & 0.8 & 0.887 & 0.998 & 0.975 \\
3 & [4, 4, 2, 9, 2, 2, 1, 1, 1, 1] & 0.742 & 0.454 & 0.825 & 0.65 \\
4 & [1, 4, 4, 2, 2, 2, 9, 1, 1, 1] & 0.285 & 0.659 & 0.688 & 0.325 \\
5 & [1, 4, 4, 2, 2, 2, 1, 1, 1, 9] & 0.285 & 0.697 & 0.667 & 0.325 \\
6 & [1, 1, 1, 1, 2, 2, 2, 4, 4, 9] & -0.8 & 0.149 & 0.571 & 0.0 \\
\hline
\end{tabular}
\caption{} 
\label{table1}
\end{table*}

\section{Experiments}\label{experiments}

In this section, we show that rankDCG satisfies all the objectives and outperforms conventional rank-ordering measures on the constructed and the real data. The specified objectives are the following:
\begin{enumerate}
  \item correctly work with multiple ties
  \item address non-normal rank value distribution
  \item emphasize correct ordering of high-rank elements
  \item produce consistent and meaningful scoring range
\end{enumerate}

\subsection{Constructed data}
We evaluate the behavior of rankDCG in seven possible scenarios: 1) perfect ordering, 2) misplacing low-rank elements,  3) misplacing a high-rank element with a medium rank element,  4) and 5) misplacing high and lows rank elements, and 6) the worst case (reversed ordering). 
All the experiments are conducted on the list L = [9, 4, 4, 2, 2, 2, 1, 1, 1, 1] defined in the previous section and hypothesis list in Table \ref{table1}. The results can be found in Table \ref{table1}.
\\

From the table, you can see that only rankDCG satisfies our criteria. Starting with the objective 1, we can see that only Kendall's $\tau$ and rankDCG address it properly. The score of comparing reference list L and lists 4 and 5 from the table produce the same score. This fact brings robustness to possible element permutations inside a subgroup of elements with the same rank. \\

The second and third objectives are the ability to deal with a non-normal distribution and emphasize correct ordering of rare, top-rank, and elements. RankDCG produces the most accurate cost function. This can be observed by comparing reference list L to lists 2, 3 and 4 in the table. In the case of the comparison with list 2, most measures produced reasonable results. NDCG puts little cost on misidentifying low-rank elements. This score follows my rankDCG, with AveP and $\tau$ being the harshest score of 0.8 for miss-ordering low-rank element. On the other hand, $\tau$ puts very little cost on misplacing the top element (0.742). This fact makes high-rank element ordering of a lesser importance. If we look at case 4, we can see that AveP gives a higher score of 0.650 for placing the top-rank element into the lowest-rank group, compare to 0.454 score for placing the same element into a better subgroup. Among all score variations, rankDCG fits right in the middle with the scoring cost function and produces a linear score decrease with worse ordering case. \\

Finally, due to the initial application of the surveyed measures in the IR area, none of the measures satisfies the lower bound requirement. This can be observed in the case 6. In case 6, the worst case ordering (reverse), all measures produce scores that are difficult to understand. The score from $\tau$ and AveP show that the results are not good, but not the worst possible case. NDCG's score can be interpreted as a good result by a person not familiar with the measure or the task. RankDCG is the only measure that produces a comprehensive worst case score. 


\subsection{Real data}

One real world problem where common measures fall short is user ranking. This task involves ranking users according to their community rank. We are looking at the Reddit website (www.reddit.com). Reddit is a website where users create posts on different topics or share resources such as pictures, videos, or links to other resources. Users can participate in discussions through creating threads of comments. Each comment can earn comment karma, which is Reddit's form of approval. We consider data from politics subreddit. We rank users from five randomly chosen subreddits that contain at least one-hundred comments. On average, each subreddit contains 129.8 users.  Using NLP algorithms, we analyze the comments and predict user rank (karma index) based on text analysis. This problem is very challenging and the results are far from perfect. However, to demonstrate shortcomings of popular rank-measures we create four tests: 1) we limit the data and produce a bad ranking prediction using limited part-of-speech analysis, 2) a slightly better rank predictions using LIWC word list \cite{pennebaker2001linguistic}, 3) further improved ranker using n-gram approach, and 4) the perfect prediction, comparing the reference with itself. The results can be found in Table \ref{table2}.\\

\begin{table}[!htbp]
\centering
\begin{tabular}{ | l | c | c | c | c |}
\hline
\# &  Kendall's $\tau$ & AveP & nDCG & rankDCG \\  
\hline\hline
1  & nan & 0.79 & 0.883 & 0.0 \\
2  & 0.197 & 0.668 & 1.188 & 0.32 \\
3  & 0.136 & 0.585 & 1.318 & 0.347 \\
4  & 0.5 & 1 & 1 & 1 \\

\hline
\end{tabular}
\caption{} 
\label{table2}
\end{table}

From the Table \ref{table2}, we can see a few interesting cases. In the first case, the rank-ordering algorithm based on limited data outputs the majority class. As a result, we can see that Kendall's $\tau$ cannot deal with this case while AveP and nDCG returned seemingly good results. In the second case, with a slightly better ranking model, all measures show improvement. nDCG returns a score higher than one because the algorithms overpredict high ranks. The third case, a better model, perceives the results as worse than the results from the second case by Kendall's $\tau$ and AveP. nDCG, on the other hand, produces a higher than 1 score. In the last fourth case, the perfect ordering, all measures, with the exception of Kendall's $\tau$, produce correct scoring. After considering all the cases, we can see that only rankDCG shows consistent evaluation scores with a gradual improvement of the algorithm and meaningful lower and upper bounds.

\section{Conclusion}\label{conclusion}

In this paper, we present rankDCG, a rank-ordering evaluation measure. RankDCG is a modification of popular nDCG algorithm that addresses some shortcomings of common evaluation measures. While there is a number of popular evaluation measures available, we show that they cannot properly evaluate ranking problems with discrete values, multiple ties, and nonlinear rank distribution. In this work, we define criteria that a good evaluation measure needs to submit and show that among popular measures only rankDCG satisfies it. We release rankDCG evaluation package to the public as a part of this work and make it available on github\footnote{https://github.com/dkaterenchuk/ranking\_measures}.


\newpage

\section{Bibliographical References}
\label{main:ref}

\bibliographystyle{lrec2016}
\bibliography{xample}


\end{document}